\documentclass[letterpaper, 10pt, twocolumn]{article}
\usepackage{usenix}

\usepackage[utf8]{inputenc}
\usepackage{url} 
\usepackage{paralist, xspace}
\usepackage{amsmath, amsfonts, amsthm}
\usepackage{graphicx}

\usepackage{xcolor}

\newcommand{\projname}{STAR-Vote\xspace}
\newcommand{\elgamal}{Elgamal\xspace}  

\title{\projname: \\
A Secure, Transparent, Auditable, and Reliable Voting
System}

\author{
Josh Benaloh (Microsoft Research) \and
Mike Byrne (Rice University) \and
Philip Kortum (Rice University) \and
Neal McBurnett (ElectionAudits) \and
Olivier Pereira (Université catholique de Louvain) \and
Philip B. Stark (University of California Berkeley) \and
Dan S. Wallach (Rice University)}

\begin{document}
\maketitle

\begin{abstract}
In her 2011 EVT/WOTE keynote, Travis County, Texas County Clerk
Dana DeBeauvoir described the qualities she wanted in her ideal election system
to replace their existing DREs.
In response, in April of 2012, the authors, working with DeBeauvoir and her staff,
jointly architected \projname, a voting system with a DRE-style human interface
and a ``belt and suspenders'' approach to verifiability.
It provides both a paper trail and end-to-end cryptography using COTS hardware.
It is designed to support both ballot-level risk-limiting audits,
and auditing by individual voters and observers.
The human interface and process flow is based on modern usability research.
This paper describes the \projname architecture, which could well be
the next-generation voting system for Travis County and perhaps
elsewhere.

{\em This paper is a working draft. Significant changes should be
  expected as the STAR-Vote effort matures.}
\end{abstract}

\section{Introduction}

\label{sec:intro}

A decade ago, DRE voting systems came with a promise of improvement. By having a
computer mediating the user's voting experience, they could ostensibly
improve usability through summary screens and a
variety of accessibility features including enlarged text, audio
output, and specialized input devices. They also promised to improve
the life of the election administrator, yielding quick, accurate tallies without
any of the ambiguities that come along with hand-marked paper ballots.
And, of course, they were promised to be secure and reliable, tested
and certified. In practice, much of this was wishful thinking.

Many current DRE voting systems experienced their biggest sales
volume following the demonstrated failures of punch card voting
systems in Florida in the 2000 presidential election. The subsequent Help America Vote Act
provided a one-time injection of funds that made these purchases
possible. Now, a decade later, these machines are near the end of
their service lifetimes. 

Last year, the election administration office of Travis County, Texas, an
early adopter of these DRE systems, concluded that current
systems on the market were inadequate for their need to replace their
end-of-life DRE systems. They were also unhappy with the
current-generation precinct-based optical scanned paper ballot systems
for a variety of reasons. In particular, hand-marked paper ballots
open the door to ambiguous voter intent, which Travis County unhappily
had to deal with in its previous centrally-tabulated optical scan
system. They didn't want to go back. Likewise, with early
voting and election day vote centers that must be able to give
any voter who arrives at
any location the proper ballot style, pre-printed paper ballots would
be a management nightmare.
Ballot-on-demand printing systems
require laser printers that cannot run all day on battery backup
systems\footnote{A laser printer may consume as much as 1000 watts
  while printing. A reasonably good UPS, weighing 26~kg, can provide
  that much power for only ten minutes. Since a printer must take time
  to warm up for each page when printed one-off (perhaps 10
  seconds total per page), as few as 60 ballots could be printed before the
  battery would be exhausted.},
which restricts their reliability.

A group of academic experts in voting systems was assembled 
to design a replacement system in sufficient
detail that bids could be solicited from manufacturers to implement
the system. Our group included experts in cryptography, auditing, and
usability, leading to some interesting challenges and questions. We
were given several basic constraints: The user experience must
resemble current DRE systems, but there should be a tangible
voter-verifiable paper ballot, printed by the machine and deposited by
the voter into a physical ballot box. This would allow for fast machine
tallies and statistical audits to ensure their equivalence to the
paper records. The system must be able to run all day on battery power.
We were free to recommend sophisticated end-to-end
cryptographic methods, privacy-preserving risk-limiting auditing
methods, and pretty much anything else we felt was beneficial, within
these constraints. Of course, reducing cost was also desirable as was
anything that might reduce the burden for poll workers or voters. The
issue of federal or state election certification, for the purpose of
this exercise, was considered out of scope. (Yes, it's a real
challenge, but it wasn't our challenge.)

\section{Voter Flow}

\label{sec:voterflow}

\begin{figure*}
\includegraphics[width=6.5in]{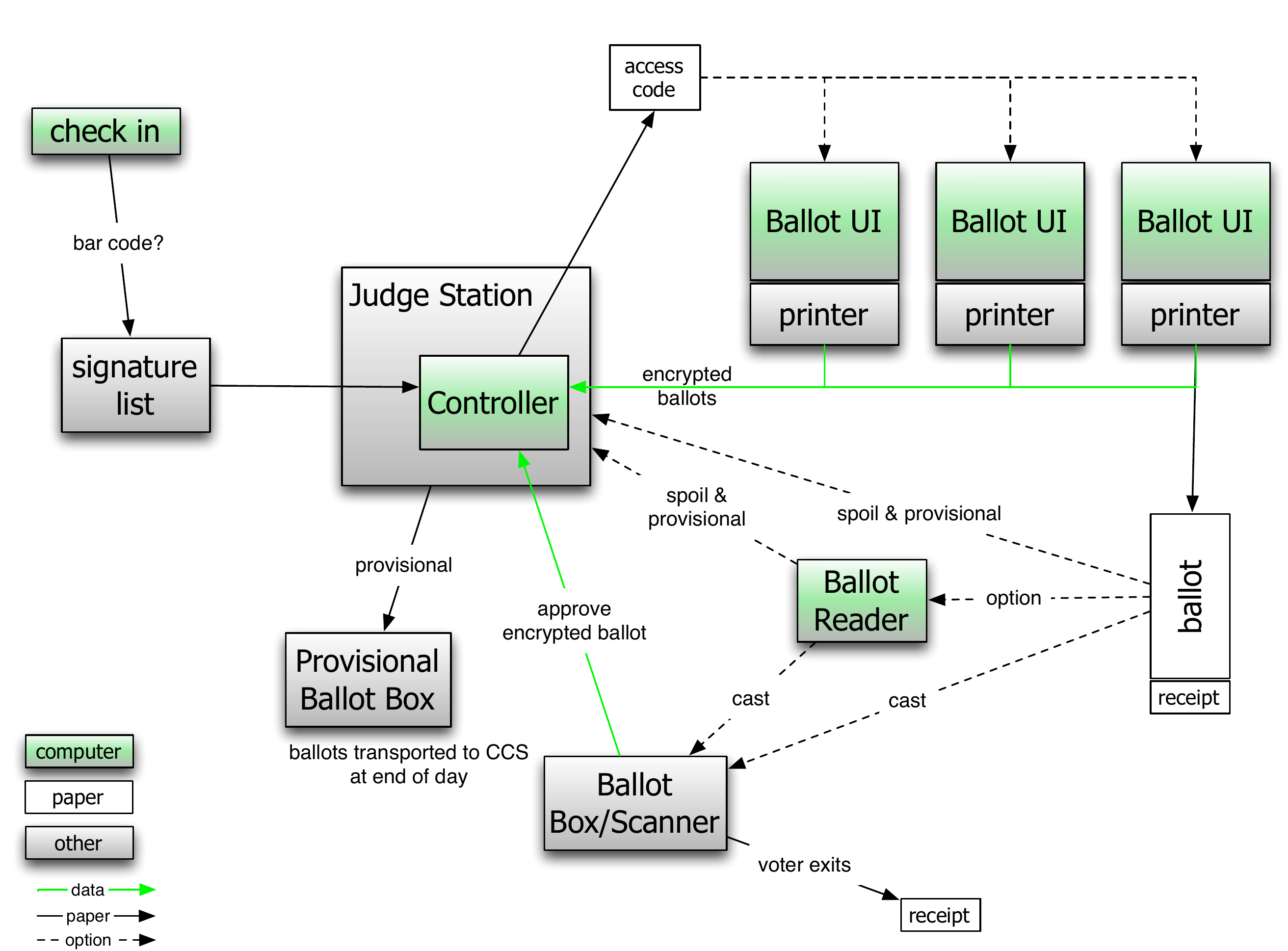}
\caption{The design of the \projname system. Green objects are computers, white objects are paper records, and other objects are shaded in gray. Arrows display the flow of information; green for digital information, black for paper, and dashed lines indicate that the flow is contingent on voter choice.\label{fig:design}.}
\end{figure*}

Figure~\ref{fig:design} shows how \projname works from the perspective
of a voter going through the system. 
The \projname voting system bears a resemblance to the Hart InterCivic eSlate system and to VoteBox~\cite{sandler08votebox}, in that the voting machines are networked together, 
simplifying the movement of data. 
Like eSlate, our design contains a networked group of voting machines that share a common judge's station with a computer like Hart InterCivic's ``Judge Booth Controller'' (JBC) that manages everything. 

\begin{enumerate}
\item {\em Check-in (pollbook).}
The first step for the voter is to check-in with a poll worker. This is where voter registration is verified and the voter's precinct is identified so that the appropriate ballot style can be generated. The voter also signs into some sort of paper book. Subsequently, the voter will receive something to take to the judge's station that identifies the proper ballot style for the voter. If the voter's registration cannot be verified, a provisional ballot will be used. Notably, {\em there will not be an electronic path between the voter registration phase and subsequent phases}.\footnote{
It would actually be possible to allow provisional voters to vote along with other voters uses the STAR sytem and tag the record for posible later removal, but for simplicity we do not include this process herein.}
Instead, the token, likely from a thermal printer, will only offer a number to identify the ballot style. This might also be encoded as a one-dimensional bar code to reduce data entry errors. Nothing on this token is secret, nor is it unique to any individual voter.

\item {\em Receive token.}
The voter takes the ballot-style identifying token to a poll worker at the judge's station which then issues another token, again probably a piece of paper with a 5-digit code on it. (There will probably need to be a special alternative for ADA compliance as not all voters can see or handle paper.) 

\item {\em Select machine.}
The voter possibly queues at this point, and then selects from one of the available voting stations.

\item  {\em Enter token.}
The first thing the voter does at the voting station is enter the digits on the token. 
This action, in turn indicates the voter's authenticity and identifies the desired ballot style. The token will also be flagged as provisional if that applies. 
Once the voter enters the code, it's transmitted over the local network to the judge's station, invalidating it immediately. Only a small number of these codes will ever be active at any time, allowing codes to be generated randomly and reused through the day. 
There will be no permanent record binding this code to the voter, as that could compromise voter anonymity.\footnote{Hart InterCivic generates comparable codes, but they're not random. Every voter's code is unique to that voter / judge station combination. \projname fixes this problem.}

\item {\em Make selections.}
 The voter makes selections on the GUI (for sighted voters)
 or auditory UI (for non-sighted voters).
 There is a review screen (or the auditory equivalent)
 so that the voter can confirm all selections before producing a paper record.

\item {\em Print.} When the voter has finished making selections,
  the voting terminal prints two (possibly joined) items:
  (1) a paper ballot which includes a human-readable summary of the voter's selections
  and a random (non-sequential) serial number, and
  (2) a paper a take-home receipt that identifies the voting terminal used, the
  time of the vote, and a short (16-20 character) hash code which
  serves as a commitment to the vote but does not reveal its contents.
  The voting terminal also sends data about the vote and receipt
  to the judge's station.\footnote{
  Specifically, the voting terminal sends the judge's station
  the raw ballot selections, an encryption of the ballot selections,
  the hash code derived from this encryption
  which is printed on the voter's take-home receipt,
  and the serial number which is printed on the paper ballot summary.}

%

\item {\em Review printed record.}
The voter may then review the printed record to confirm the indicated selections. There will be at least one station available that can OCR the paper record and read it back to the voter for those who cannot visually read the paper record.

\item {\em Option: Cast or challenge/spoil.}
After reviewing the ballot, the voter has a choice: Cast the ballot or spoil it.
 A voter might spoil the ballot because of an error (or change of heart)
 or because the voter wishes to challenge the voting terminal, demanding it to show
 that the voter's selections were correctly recorded and committed to.\footnote{
 This follows the cryptographic verification process introduced by Benaloh
 in \cite{benaloh06simple} and \cite{benaloh07evt}.}
 The two procedures are described below. Note also that there is a special procedure for provisional ballots.

 Regardless of which case, the voter may keep the take-home paper receipt.
 We note that most thermal printers include a cutting device that leaves a
 small paper connection between the two sides of the cut.
 It is therefore a simple matter for the voting terminal to print a single
 sheet that the voter can easily separate into the ballot summary and the
 take-home receipt.
 We also note that ``privacy sleeves'' (i.e., simple paper folders) can protect the privacy of these printed ballots as voters carry them from the voting machine to either the ballot box, to be cast, or the judge's station, to be spoiled.

\begin{enumerate}
\item  {\em Cast ballot.}
A voter who wishes to cast the ballot takes the paper ballot summary to the ballot box.
The ballot box has a simple scanner that can read the serial number from the ballot
(the serial number might also be represented as a one-dimensional barcode for reliability)
and communicate this to the judge's station, allowing the judge's station to keep a record of which ballots have found their way to the ballot box, and thus, which ballots should be tabulated. 
{\em An electronic ballot record is not considered complete and should not be
 included in the tally unless and until its corresponding paper ballot summary has
 been deposited in the balot box.}

\item {\em Spoil ballot.}
If the paper record is to be spoiled, the voter returns to a poll worker at a judge's station. 
The ballot serial number is scanned so that the judge's station can record that the ballot is to be spoiled. Furthermore, the judge's station knows that the corresponding encrypted ballot record should never be used in a real tally. 
Instead, it should be decrypted and published as such. 
The original printed paper ballot thus corresponds to a {\em commitment\/} by the voting machine, before it ever knew it might be challenged.
 If the voting machine cannot produce a suitable proof that the ballot encryption matches the plaintext,
 then it's been caught cheating.
 Of course, for voters who don't care about verification,
 they can simply restart the process.
 For voters who may feel uncomfortable with this process,
 as it might reveal their intent to a poll worker,
 we note that voters could deliberately spoil ballots that misstate their true intent. We note that dedicated election monitors could be allowed to use voting machines, producing printed ballots that they would be forbidden from placing in the ballot box, but which would be spoiled and then the corresponding ciphertext would be decrypted. In effect, election monitors can conduct {\em parallel testing in the field\/} on any voting machine at any time during the live election.

\item {\em Provisional ballot.}
In the case of a provisional ballot, the voter does not have the cast vs.~spoil option, and must return the ballot to a poll worker, who places it into a distinct provisional ballot box. The voter may retain the receipt to see if the ballot ends up being counted.
\end{enumerate}

\item {\em At home (optional): Voter checks crypto.}
The encrypted votes will be posted on a public ``bulletin board''
 (i.e., a web site maintained by the county).
 The voter receipt corresponds to a cryptographic hash of the encrypted vote.
 A voter should be able to easily verify that this vote is present on the bulletin board.
 If a voter spoiled a ballot, that should also be visible on the bulletin board
 together with its decrypted selections.  This allows independent observers to
 know which ballots to include in the tally and allows independent verifiers
 to check that all spoiled ballots are correctly decrypted.  Individual voters can
 check, without any mathematics,
 that the decryptions of their own spoiled ballots match their expectations.
\end{enumerate}



\section{Design}

\label{sec:design}

From the perspective of voters, the process of registration and poll-station sign-in is unchanged from current practice.  Once authorized, voters proceed to a voting terminal where they use rich interface that prevents overvotes, warns of undervotes, and supports alternative input/output media for disabled and impaired voters. The printed ballot summary, as well as the corresponding electronic ballot record, includes a variety of cryptographic features, which we now describe.




\subsection{Crypto Overview}
From the perspective of election officials, the first new element in the election regimen is 
to generate the cryptographic keys.  
A set of election trustees is designated as key holders and a threshold number is fixed.  
The functional effect is that if there are $n$ election trustees and the threshold value is $k$, then any $k$ of the $n$ trustees can complete the election, even if the remaining $n-k$ are unavailable.  This threshold mechanism provides robustness while preventing any fewer than $k$ of the trustees from performing election functions that might compromise voter privacy. 
Threshold cryptosystems are straightforward extensions of traditional public-key cryptosystems~\cite{desmedt90threshold}.

The trustees each generate a key pair consisting of a private key and
a public key; they publish their public keys.  
A standard public
procedure is then used to compute a single public key from the $n$
trustee public keys such that decryptions can be performed by any $k$
of the trustees.  This single election public key $K$ is published and
provided to all voting terminals together with all necessary
ballot style information to be used in the election.  
Each voting terminal is also seeded with a start value $z_0$ that includes a
unique identifier for both the voting office and the election.

During the election, voters use voting terminals to make their
selections.  Once selections are completed, the voting terminal
produces paper printouts of two items.  The first is the paper ballot summary
which consists of the selections made by the voter and also includes a
random (non-sequential) serial number.  The second is a receipt that
consists of an identification number for the voting terminal,
the date and time of the vote, and a short (16-20 character) hash of the encryption of
the voter's selections together with the previous hash value.
Specifically, if the voter's selections are denoted by $v$, the
$i^{{\mathrm th}}$ hash value produced by a particular voting terminal $m$ in an election is computed as
\[
z_i=H(E_K(v),m,z_{i-1} )
\]
\noindent
where $H$ denotes the hash function and $E$ denotes encryption.

The voting terminal should retain both the encrypted ballots and the current hash value.  At the conclusion of the election (if not sooner), the encrypted ballots should be posted on a publicly-accessible web page and digitally signed by the election office using a simple signature key (not the key generated by the trustees).  The posting of each encrypted ballot should also include a non-interactive zero-knowledge (NIZK) proof that the ballot is well-formed.
Once they receive their ballots summaries and take-home receipts, voters may either deposit their ballot summaries into a ballot box or take them to a poll-worker and have them spoiled.  Ballot summaries deposited in a ballot box have their serial numbers scanned and recorded.  The electronically stored encrypted vote is not considered complete (and not included in the tally) unless and until its corresponding serial number has been recorded in the ballot box.

Any electronic stored encrypted ballots for which no corresponding serial number has been scanned and recorded are deemed spoiled.  The published election record should include all spoiled ballots as well as all cast ballots, but for each spoiled ballot the published record should also include a verifiable decryption of the ballot's contents.  Voters should be able to easily look up digitally-signed records for any receipts they hold and verify their presence and, for spoiled receipts, the ballot contents.

A voter who takes a completed paper ballot summary to a poll worker can request that the poll worker spoil the ballot and give the voter an opportunity to re-vote.  
The poll worker marks both the take-home receipt and the paper ballot summary as spoiled (including removing or marking the serial number so that it will not be recorded if subsequently placed in the ballot box) and returns the spoiled ballot summary to the voter.

Upon completion of the election, the election office homomorphically combines the cast ballots into an aggregate encryption of the election tally (this can be as simple as a multiplication of the public encrypted ballots).  At least $k$ of the election trustees then each perform their share of the decryption of the aggregate as well as individual decryptions of each of the spoiled ballots.  The trustees also post data necessary to allow observers to verify the accuracy of the decryptions.

A privacy-preserving risk-limiting audit is then performed by randomly selecting paper ballot summaries and matching each selected ballot with a corresponding encrypted ballot to demonstrate the correct matching and provide software-independent evidence of the outcome~\cite{rivest06sivoting,lindemanStark12,starkWagner12}.

\subsection{Triple Assurance}

Three lines of evidence are produced to support each election outcome~\cite{starkWagner12}.  The homomorphic tallying process proves that the announced tally corresponds to the posted encrypted ballot records.  The ballot challenge and receipt checking processes allow voters to check that these encrypted ballot records correctly reflect their selections.  The risk-limiting audit process provides an independent check that a hand count of the paper ballots matches the outcome which a hand count of the paper records would produce.  In addition, the 
paper records remain available in case of systemic failure of the electronic records or 
if a manual count is ever desired.
The paper and electronic records are conveyed to the local election office separately, providing
additional physical security of the redundant audit trail.

The design of the election system ensures that all three of these checks should be perfectly consistent.  There is sufficient information in the records so that if any discrepancies arise (for instance because of loss of some of the electronic or paper records), the discrepancies can be isolated to individual ballots that are mismatched or counted differently.

\subsection{Software and Hardware Engineering}

An important criteria for \projname is that it should leverage commodity components whenever feasible. This reduces cost and simplifies the ability for an election administrator to replace aging hardware by sourcing it from multiple vendors. While this paper isn't intended to cover certification issues, the separation of hardware and software allows for the possibility of {\em commercial off-the-shelf} (COTS) hardware, which {\em could} be subject to a lower bar for certification than the software.

Ideally, the voting terminals and the judge station could use identical hardware. In particular, we believe that a reasonable target might be ``point of sale'' terminals. These are used in restaurants worldwide. They are used in relatively demanding environments and, on the inside, are ordinary PCs, sometimes built from low-power laptop-class parts. The only missing hardware from a COTS point of sale terminal, relative to our needs for \projname, are a printer and a battery.

If you want a reliable, low-power printer, without having to worry about consumable ink or toner, there's only one choice: thermal printers. They come in a variety of widths, up to US Letter size. Thermal paper, particularly higher cost thermal paper, can last for years in an air-conditioned warehouse, although some experimentation would be required to see whether it can survive an un-air-conditioned trip in a hot car in the summer. Every shipping label from major online vendors like Amazon is printed thermally, lending some credence to its survivability in tough conditions.

Specifying a battery is more complicated. We could require that the voting terminal have an internal (and removable) battery, but this eliminates COTS point of sale terminals. Tablet computers come with built-in batteries that, at least in some cases, can last all day. Tablet computers have smaller screens than we might prefer, and we would have to worry about theft. Also, we would prefer to use wired networks, rather than the wireless networks built into most tablets. Perhaps the right answer is to specify a point of sale terminal with an external battery, and hope a vendor can do this without extensive customization.

For the software layer, we see no need for anything other than a commodity operating system, like Linux, which can be stripped of unessential features to reduce the attack surface. For example, we don't need a full-blown window system or 3D graphics pipeline. All we need are basic pre-rendered ballots, as in pVote~\cite{yee06prui,yee07pvote} or VoteBox~\cite{sandler08votebox}. We would specify that the voting system software be engineered in a type-safe language like Java or C\# (eliminating buffer overflow vulnerabilities, among other problems) and we would also specify that the software be engineered with {\em privilege separation}~\cite{PFH03}, running separate parts of the voting software as distinct applications, with distinct Unix user-ids, and with suitably reduced privileges. For example, the storage subsystem can maintain append-only storage for ballots. The voter-facing UI would then have no direct access to ballot storage, or the network, and could be ``rebooted'' for every voter. Consequently, a software compromise that impacts the UI application could impact at most one voter.

A separation architecture like this also provides some degree of protection over sensitive cryptographic key materials, e.g., if we want every voting terminal to have a unique private key to compute digital signatures over ballots, then we must restrict the ability for compromised software to extract the private keys. DStar~\cite{dstar2008}, for example, used this technique to protect the key material in an SSL/TLS web server.

\section{Usability}

\subsection{Design Considerations}
In designing this reference voting system it was important to maximize the usability of the system within the framework of enhanced security and administrative expediency. The overall design of the system was strongly influenced by usability concerns. For example, a proposal was put forth to have all voters electronically review the paper record on a second station; this was rejected on usability grounds.
ISO 9241 Part 11~\cite{iso1998} specifies the three metrics of usability as effectiveness, efficiency, and satisfaction, and these are the parameters we attempt to maximize in this design. Effectiveness of the system means that users should be able to reliably accomplish their task, as they see it. In voting, this means completing a ballot that correctly records the candidate selections of their choice, whether that be though individual candidate selection by race, straight party voting, or candidate write–ins. Efficiency measures the ability of a voter to complete the task with a minimum of effort, as measured through time on task or number of discrete operations required to complete a task. Efficiency is important because users want to complete the voting task in the minimum amount of time and voting officials are concerned about voter throughput. Reduced efficiency means longer lines for waiting voters, more time in the polling booth, and more equipment costs for election officials. Satisfaction describes a user's subjective assessment of the overall experience. While satisfaction does not directly impact a voter's ability to cast a vote in the current election, it can have direct impact on their willingness to engage in the process of voting at all, so low satisfaction might disenfranchise voters even if they can cast their ballots effectively and efficiently. 
How does this design seek to maximize these usability metrics? For voting systems, the system must generally be assumed to be walk-up-and-use. Voting is an infrequent activity for most, so the system must be intuitive enough that little to no instruction is required to use. The system should minimize the cognitive load on voters, so that they can focus on making candidate selections and not on system navigation or operation. The system should also mitigate the kinds of error that humans are known to make, and support the easy identification and simple correction of those errors before the ballot is cast. 
\paragraph{Why not paper?}
Paper ballots (bubble ballots in particular) exhibit many positive characteristics that make them highly usable~\cite{hfes-06,byrne-baseline}. Users are familiar with paper, and most have had some experience with bubble-type item selection schemes. Voting for write-in candidates is simple and intuitive. Unlike electric voting machines, paper is nearly 100\% reliable and is immune from issues of power interruption. Further, paper leaves an auditable trail, and wholesale tampering is extremely difficult. 
For all these benefits, paper is not the perfect solution. Voters actually show higher satisfaction with electronic voting methods than they do with paper~\cite{everett08chi-dre-usability} and paper still has some significant weaknesses that computers can overcome more easily. 
First, the ambiguity that can be caused by partial marks leads to substantial problems in counting, recounting, and re-intrpeting paper ballots.
Second, voting by individuals with disabilities can be more easily accommodated using electronic voting methods (e.g., screen readers, jelly switches).
Third, electronic voting can significantly aid in the reduction of error (e.g. undervotes, overvotes, stray marks) by the user in the voting process.
Forth, electronic voting can more easily support users whose first language is not English, since additional ballots for every possible language request do not have to be printed, distributed and maintained at every polling location. This advantage is also evident in early voting and vote center administration; rather than having to print, transport, secure, and administer every possible ballot for every precinct, the correct ballot can simply be displayed for each voter. The use of computers also allows for the inclusion of sophisticated security and cryptography measures that are more difficult to implement in a pure paper format.
Finally, administration of the ballots can be easier with electronic formats, since vote counting and transportation of the results are more efficient.
We have taken a hybrid approach in this design, by using both paper and electronic voting methods in order to create a voting system that retains the benefits of each medium while minimizing their weaknesses. 
\paragraph{Usability vs Security}
Usability and security are often at odds with each other. Password design is a perfect example of these opposing forces. A system that requires a user have a
32-character password with upper and lower case letters, digits, and symbols with no identifiable words imbedded might be highly secure, but it would have significant usability issues. Further, security might actually be compromised as users would be likely to write the password down and leave it in an insecure location (like the computer monitor). In voting we must strive for maximum usability while not sacrificing the security of the system (our security colleagues might argue that we need to maximize security while not sacrificing usability). In our implementation, many of the security mechanisms are invisible to the user. Those that are not invisible are designed in such a way that only those users who choose to exercise the enhanced security/verifiability of the voting process are required to navigate additional tasks (e.g., ballot challenge, post-voting verification). 
\paragraph{Error reduction}
The use of computers in combination with paper is anticipated to reduce errors committed by voters. Because voters will fill out the ballot on electronic voting terminals, certain classes of errors are completely eliminated. For example, it will be impossible to over vote or make stray ballot marks, as the interface will preclude the selection of more than a single candidate per race. Under voting will be minimized by employing sequential race presentation, forcing the voter to make a conscious choice to skip a race~\cite{greene-thesis}. Undervotes will also be highlighted in color on the review screen, providing further opportunity for users to reduce undervotes. This review screen will also employ a novel party identification marker (described below) that will allow a voter to easily discern the party for which they cast a vote in each race. The use of the paper ballot (printed when the voter signals completion) provides the voter with a final additional chance to review all choices before casting the final ballot. 

\subsection{User Interface Design Specification}
The basic design for the UI is a standard touchscreen DRE with auditory interface for visually impaired voters and support for voter-supplied hardware controls for physical impairments (e.g., jelly switches).
\paragraph{The VVSG}
The starting point for UI specifications is the 2007 version of the Voluntary Voting System Guidelines (VVSG). These guidelines specify many of the critical properties required for a high-quality voting system user interface, from simple visual properties such as font size and display contrast to more subtle properties such as ballot layout. They also require that interfaces meet certain usability benchmarks in terms of error rates and ballot completion time. We believe that no extant commercial voting UI meets these requirements, and that any new system that could meet them would be a marked improvement in terms of usability. That said, there are some additional requirements that we believe should be met. 
\paragraph{Accessibility}
While the VVSG includes many guidelines regarding accessibility, more recent research aimed at meeting the needs of visually-impaired voters~\cite{piner-11} has produced some additional recommendations that should be followed. These include:
\begin{itemize}
\item  The system should include an auditory interface than can be used either alone or in conjunction with the visual interface. 
\item Speech rate (as well as volume) should be adjustable by the voter. 
\item In order to maximize intelligibility, a synthesized male voice should be used so that speed can be altered without changing pitch. 
\item Navigation should allow users to skip through sections of speech that are not important to them as well as allowing them to replay any parts they may have missed or not comprehended the first time.
\item At the end of the voting process, a review of the ballot must be included, but should not be required for the voter. 
\end{itemize}
\paragraph{Review Screens}
Another area where the VVSG can be augmented concerns review screens. Voter detection of errors (or possible malfeasance) on review screens is poor~\cite{everett07thesis}, but there is some evidence that UI manipulations can improve detection in some cases~\cite{campbell-evt09}. Thus, \projname  requires the following in addition to the requirements listed in the VVSG:
\begin{itemize}
\item  Full names of contests and candidates should be displayed on the review screen; that is, names should be text-wrapped rather than cut off. Party affiliation should also be displayed.
\item Undervotes should be highlighted using an orange-colored background. 
\item Activating (that is, touching on the visual screen or selecting the relevant option in the auditory interface) should return the voter to the full UI for the selected contest.
\item In addition to party affiliation in text form, graphic markings should be used to indicate the state of each race: voted Republican, voted Democratic, voted Green, etc.---with a distinctive graphic for ``not voted'' as well. These graphic markings (perhaps a donkey for the Democratic Party, an elephant for the Republican Party, etc.) should be highly distinguishable from each other so that a rapid visual scan quickly reveals the state of each race.
\end{itemize}

\paragraph{Paper Record}
The VVSG has few recommendations for the paper record. For usability, the paper record should meet the VVSG guidelines for font size and should contain full names for office and candidate. To enable scanner based recounts (if necessary), the font used should be OCR-friendly. Contest names should be left-justified while candidate names should be right-justified to a margin that allows for printing of the same graphic symbols used in the review screen to facilitate rapid scanning of ballots for anomalies. Candidate names should not be placed on the same line of text as the contest name and a thin horizontal dividing line should appear between each office and the next in order to minimize possible visual confusion.

\subsection{Issues that still need to be addressed}
There are still several issues that need to be addressed in order to make the system have the highest usability. The first of these is straight party voting (SPV). SPV can be quite difficult for a voter to understand and accomplish without error, particularly if voters intend to cross-vote in one or more races~\cite{campbell-ieee}. Both paper and electronic methods suffer from these difficulties, and the optimum method of implementation will require additional research. Races in which voters are required to select more than one candidate ($k$ of $n$ races) also create some unique user difficulties, and solutions to those problems are not yet well understood.


\section{Audit}

\label{sec:audit}

The E2E feature of \projname enables individual voters to confirm that their votes were included in the
tabulation, and that the encrypted votes were added correctly.
The challenge feature, if used by enough voters, assures that the encryption was honest
and that substantially all the votes are included in the tabulation.
But there might not be many voters who challenge the system; the voters who do are hardly
representative of the voting public; and some problems may go unnoticed.

The paper audit trail enables an entirely independent check that the votes were tabulated accurately, that
the visible trace of voter intent as reflected in the ballot agrees with the encryption, and, importantly, that
the winners reported by the voting system are the winners that a full hand count of the audit trail would
reveal.
The key is to audit the machine interpretation against a manual interpretation of the paper ballots,
using a risk-limiting method.
\projname uses SOBA \cite{benalohEtal11} for this purpose.

A risk-limiting audit guarantees a large minimum chance of a full hand count of the audit trail if the
reported outcome (i.e., the set of winners) disagrees with the outcome that the full hand count would reveal.
The full hand count then sets the record straight, correcting the outcome before it becomes official.
Risk-limiting audits are widely considered best practice for election audits \cite{bestPractices08}.

The most efficient risk-limiting audits, ballot-level comparison audits, rely on comparing 
the machine interpretation of individual ballots
(cast vote records or CVRs) 
against a hand interpretation of the same ballots \cite{stark10d,benalohEtal11,lindemanStark12}.
Current federally certified voting systems do not report cast vote records, so they cannot
be audited using the most efficient techniques \cite{lindemanStark12,starkWagner12}.
This necessitates expensive work-arounds.\footnote{%
    For instance, a {\em transitive audit\/} might require marking the ballots with unique identifiers
    or keeping them in a prescribed order, re-scanning all the ballots to make digital images,
    and processing those images with software that can construct CVRs from the images and
   associate the CVRs with the ballots.
   That software in turn needs to be programmed with the all the ballot definitions in the contest,
    which itself entails a great deal of error-prone handwork.
}
The preamble to conducting a ballot-level comparison audit using currently deployed voting systems
can annihilate the efficiency advantage of ballot-level comparison
audits \cite{starkWagner12}.

A big advantage of \projname is that it records and stores individual cast vote records in a way that 
{\em can\/} be linked to
the paper ballot each purports to represent, through the encrypted vote data.
This makes ballot-level comparison audits extremely simple and efficient.
It also reduces the vulnerability of the audit to human error, such as accidental changes to the order
of the physical ballots.\footnote{%
   For instance, we have seen groups of ballots dropped on the floor accidentally;
   even though none was lost, restoring them to their original order was impossible.
}

A comparison audit can be thought of as consisting of two parts:
Checking the addition of the data,\footnote{%
   This presupposes that the contest under audit is a plurality, majority, super-majority, or vote-for-$k$
   contest.
   The operation that must be checked to audit an instant-runoff contest is not addition, but the
   same principle applies.
}
and randomly spot-checking the accuracy of the data added, to confirm that they are accurate
enough for their sum to give the correct electoral outcome.
The data are the votes as reported by the voting system.
For the audit to be meaningful, the election official must commit to the vote data before the
pot-checking begins.
Moreover, for the public to verify readily that the reported votes sum to the reported contest totals,
it helps to publish the individual reported votes.
However, if these votes were published ballot by ballot, pattern voting could be used to signal voter identity,
opening a communication channel that might enable 
widespread wholesale coercion~\cite{rescorla09,benalohEtal11}.

The SOBA risk-limiting protocol \cite{benalohEtal11} solves both of these problems:
It allows the election official to commit cryptographically and publicly to the vote data; it publishes
the vote data in plain text but ``unbundled'' into separate contests so that pattern voting cannot
be used to signal.
Moreover, the computations that SOBA requires are extremely simple (and could be simplified
even further using the results in \cite{lindemanStark12}).
The simplicity increases transparency, because observers can confirm that the calculations
were done correctly, using a pencil and paper or a hand calculator.

The encrypted vote data on the ballot that \projname produces serves as a unique ballot identifier,
an ingredient that simplifies SOBA procedures because it eliminates the need to store ballots in a rigid order.
Moreover, because the voting terminal generates both the electronic vote data and the paper ballot, 
the audit should find very few if any discrepancies between them.

But since voters and election workers will handle the ballots in transit from the voting terminal 
to the scanner to the audit, voters might make marks on their ballots.
Depending on the rules in place for ascertaining voter intent from the ballot,
those marks might be interpreted as expressing voter intent different from the
printed selections, in which case the SOBA audit might find discrepancies.

It could also happen that a ballot enters the ballot box but its serial number is not
picked up, so the electronic vote data ends up in the ``untallied but unspoiled'' group.
This should be detectable by a compliance audit \cite{benalohEtal11,lindemanStark12,starkWagner12} 
as a mismatch between the number of recorded votes and the number of pieces of paper,
providing an opportunity to resolve the problem before the audit begins.

If such cases remain and turn up in the audit sample, SOBA would count them as discrepancies 
and the sample would expand accordingly, either
until there is strong evidence that the electoral outcomes are correct despite any errors the audit
uncovers, or until there has been a complete hand count.

The random selection of ballots for the SOBA audit should involve public participation in
generating many bits of entropy to seed a high-quality, public, pseudo-random 
number generator (PRNG), which is then used to select a sequence of
ballots to inspect manually~\cite{lindemanStark12}.
(For instance, audit observers might roll 10-sided dice repeatedly to generate a 20-digit number.)
Publishing the PRNG algorithm adds transparency by allowing observers to verify
that the selection of ballots was fair.



\section{The Cryptographic Workflow}

\label{sec:crypto}


\paragraph{The core elements}
\label{sec:crypto-core}


The most important cryptographic element in STAR-Vote is the cryptosystem
that is used to encrypt ballots.  This should be a threshold cryptosystem
(so that decryption capabilities are distributed to protect voter privacy)
that has an additive homomorphic property (to allow individual encrypted ballots to
be combined into an aggregate encryption of the tally).  An exponential
version of the \elgamal cryptosystem satisfies the required properties.

Cryptographic key generation can be accomplished in one of two ways, depending on the availability of the election trustees and the desired
amount of robustness.  The preferred approach is a two-step process,
but a simpler one-step process can be used if the robustness is eliminated.
At the end of the key generation procedure, the trustees each hold a
private key share that does not contain any information on the full
private key, and the unique public key $K$ corresponding to those
shares is published.

During the polling phase, the ballot marking devices encrypt the votes
of each voter using the public key $K$. This encryption procedure is
randomized in order to make sure that two votes for the same
candidates result in ciphertexts that look independent to any
observer. 

A short hash value of each ciphertext is also computed, e.g., by
truncating the output of the SHA-256 hash function.  This hash provides
a unique fingerprint of the ballot, which is provided to the voter as
part of the take-home receipt. All the ciphertexts and hashes that are
computed are posted on a publicly accessible web page, either
immediately or as soon as the polls are closed. This web page is
digitally signed by the election office using a traditional signature
key (not the key generated by the trustees).

The posting of all of the encrypted ballots and hashes gives all voters
the ability to verify that their ballots have been properly recorded.
Additionally, this web page makes it possible for observers
to confirm the homomorphic aggregation of the individual ballots into a
single encryption of the sum of the ballots (which constitutes an encryption
of the election tallies). 

At the end of the election, any set of trustees that achieve the pre-set quorum
threshold use their respective private keys to decrypt the derived aggregate tally
encryption.  This procedure is simple and efficient and can be completed
without interaction between the trustees.  We note that the individual 
encrypted ballots, from which the aggregate encryption of the tallies
is formed, are never individually encrypted.  However, each spoiled ballot {\em is}
individually decrypted using exactly the same process that is used to decrypt
the aggregate tally encryption.

The elements we just described make the core of the workflow and are
sufficient to compute an election tally while preserving the privacy
of the votes. We now explain various ways in which this simple
workflow is hardened in order to make sure that the tally is also
correct. All the techniques that follow enable the verification of
different aspects of the ballot preparation and casting.

\paragraph{Hardening encryption}
\label{sec:hardening-encryption}
Since the tally does not involve the decryption of any individual
ballot, and since the audit procedure relies on the fact that all
counted ballots are properly formed, it is crucial to make sure that
all the encrypted ballots that are added correspond to valid votes.
This is achieved by requiring the ballot marking devices to compute,
together with the encryption of the votes, a non-interactive
zero-knowledge (NIZK) proof that each ballot is well-formed. Such a
proof guarantees that each ciphertext encrypts a valid vote and does not
leak any other information about the content of the vote. As a side
benefit, this proof can be designed to make the ballots non-malleable,
which provides an easy technique to prevent the replay of old ballots
(i.e., reject duplicates).

\paragraph{Hardening decryption}
\label{sec:hardening-decryption}
Making sure that the encrypted ballots are valid is not enough: we
also need to make sure that the tally is correctly decrypted as a
function of those encrypted ballots: otherwise, malicious trustees (or
trustees using corrupted devices) could publish an outcome that does
not correspond to these ballots. As a result, we require the trustees
to provide evidence of the correctness of the decryption operations
that they perform.  This can also be accomplished with NIZK proofs,
although exponential \elgamal and many other suitable cryptosystems
allow a single value to be published to enable observers to verify that
the decryption is correct.

\paragraph{Hardening the timeline}
\label{sec:hard-timeline}

The procedures described above prevent malfunctioning or corrupted
voting terminals or trustees to falsify individual ballots or decryption
operations.

The detection of manipulation of encrypted ballots can be more
effective by linking ballots with each other, using hash chaining.
For this purpose, each ballot marking device is seeded, at the beginning
of the election, with a public start value $z_0$ that includes a
unique identifier for the election.
This unique identifier should be chosen at random shortly before the
election.


This $z_0$ seed is then used as follows: as soon as a voting terminal
with identifier $m$ computes an encrypted ballot $b_i$, it
computes a hash $z_i := H(b_i \| m \| z_{i-1})$ and transmits the value
to the judge's station. A truncated version of $z_i$ forms the hash code
that appears on the voter's take-home receipt.  When the polls close,
the final $z$ value is digitally signed and made public.

As a result of this procedure, any removed ballot will
invalidate the hash chain which is committed to at the close of the
election and whose constituents appear on voter's taken-home receipts.

\paragraph{Hardening the link between the paper and electronic
  election outcome}
\label{sec:hard-link-betw}

The voting terminals print human-readable versions of each
ballot summary which can be inspected for correctness by voters.
In addition to the cast or challenge procedure discussed above,
a risk-limiting audit provides further insurance that the
election outcome that could be computed from the paper ballots matches
the one that is computed from the decryption of the encrypted election
outcome.
To support the risk-limiting audit, a cryptographic hash structure
must be built according to the SOBA schema (see section~\ref{sec:audit}).
This structure is composed from the raw votes on each of the ballots.
While it would be possible to obtain these raw votes by decrypting the
individual encrypted ballots, this operation (which would require participation
of the election trustees) is not necessary.  Instead, the voting terminals
have all the data needed to construct the SOBA structure and they can either
build the structure in real time during the election or encrypt it for
later processing.\footnote{
The encryption here is simply an operational safeguard to avoid storing raw
cast vote records on the voting terminals or judge's stations.
This can be accomplished with conventional encryption techniques and need
not involve the election trustees.}

\paragraph{The full cryptographic protocol}
\label{sec:full-protocol}
The resulting cryptographic workflow is as follows. 
\begin{enumerate}
\item The trustees jointly generate a threshold public key/private key
  encryption pair. The encryption key $K$ is published.
\item Each voting terminal is initialized with the ballot/election
  parameters, the public key $K$ and a unique seed $z_0$ that is
  computed by hashing all election parameters and using a public
  random salt.
\item When a voter completes the ballot marking process selection
  to product a ballot $v$, the voting terminal performs the following operations: 
  \begin{enumerate}
  \item It selects a unique ballot serial number $s$.
  \item It computes an encryption $c_v = E_K(v)$ of the vote, as
    well as a NIZK proof $p_v$ that $c_v$ is a valid ballot encryption.
  \item It computes a hash code $z_i = H(c_v \| p_v \| m \| z_{i-1})$,
    where $m$ is the voting terminal unique identifier.
  \item It prints a paper ballot in two parts. The first contains $v$
    in a human readable format as well as $s$ in a robust machine
    readable format (e.g., a barcode). The second is a voter take-home
    receipt that includes, the voting terminal identifier $m$, the date
    and time, and the hash code $z_i$ (or a truncation thereof).
  \item It transmits $(c_v, p_v, m, z_i, s)$ to
    the judge's station which will ultimately publish all of these values
    {\em except\/} $s$.
  \end{enumerate}
\item When a ballot is cast, the serial number $s$ is scanned and sent
  to the judge's station.  The judge's station then marks the associated ballot
  as complete and ready to be included in the tally.
\item When the polls are closed, the tally is computed: the product
  $c$ of all flagged encrypted votes is computed and verifiably
  decrypted, providing an election result $r$.
\item A hash structure is computed to support the SOBA audit described in
  section~\ref{sec:audit}.
\end{enumerate}

All stored data can then be digitally signed and published by the
local authority. Those audit data are considered to be valid if the
hash chain checks and if all cryptographic proofs check, that is, if
the ballot validity proofs check, it $c$ is computed and decrypted
correctly, and if all spoiled ballots are decrypted correctly..

\paragraph{Write-in votes}
So far, we have not described how our cryptographic construction can
support write-in voting. Support for write-in votes is required in Texas
and many other states. To be general-purpose, \projname should adopt
the vector-ballot approach~\cite{kiayias04vectorBallot}, wherein there
is a separate homomorphic counter for the write-in slot plus
an extended NIZK proof to ensure that the write-in slot only contains
a string when the write-in counter is non-zero. If there are enough
write-in votes to influence the election outcome, then the write-in slots, across the whole
election, will be mixed and tallied.

We note that, at least in Texas, write-in candidates must be
registered in advance. It's conceivable that we could simply allocate
a separate homomorphic counter for each registered candidate and have
the \projname terminal help the voter select the desired ``write-in''
candidate. Such an approach could have significant usability benefits
but might run afoul of regulations.

\section{Coercion}

\label{sec:coercion}

In designing \projname, we made several explicit decisions regarding how much to complicate the protocol and impede the voter experience in order to mitigate known coercion threats.  Specifically, one known threat is that a voter is instructed to create a ballot in a particular way but to then execute a decision to cast or spoil the ballot according to some stimulus received after the ballot has been completed and the receipt has been generated.  The stimulus could come, for example, from subtle motions by a coercer in the poll site, the vibration of a cell phone in silent mode, or some of the (unpredictable) data that is printed on the voter’s receipt.  Some prior protocols have required that the receipt, although committed to by the voting device, not be visible to the voter until after a cast or spoil decision has been made (perhaps by printing the receipt face down behind a glass barrier) and configuring poll sites so that voters cannot see or be seen by members of the public until after they have completed all steps.  We could insist on similar measures here, but in an era where cell phones with video recording capabilities are ubiquitous and eyeglasses with embedded video cameras can easily be purchased, it seems unwise to {\em require/} elaborate measures which mitigate some coercion threats but leave others unaddressed.

A similar threat of ``chain voting'' is possible with this system wherein a voter early in the day is instructed to neither cast nor spoil a ballot but to instead leave the poll site with a printed ballot completed in a specified way.  This completed ballot is delivered to a coercer who will then give this ballot to the next voter with instructions to cast the ballot and return with a new printed ballot---again completed as specified.  Chain voting can be mitigated by instituting time-outs which automatically spoil ballots that have not been cast within a fixed period after their production and by attempting to prevent voters from leaving poll sites with printed ballots, but, beyond simple mitigations, we require no additional steps to make chain voting impossible. 

(Traditional paper ballots sometimes include a perforated header section which includes a serial number. A poll worker keeps one copy of this number and verifies that the ballot a voter wishes to cast matches the expected serial number. If so, the serial number is then detached from the ballot and deposited in the box. \projname could support this, but we believe it would damage \projname's usability. The timeout mechanism seems like an adequate mitigation.)

We do, however, take measures to prevent wholesale coercion attacks such as those that may be enabled by pattern voting.  For instance, The SOBA audit process is explicitly designed to prevent pattern-voting attacks; and the high assurances in the accuracy of the tally are acheived without ever publishing the full set of raw ballots.


An interesting concern is that our paper ballots have data on them to connect them to electronic ballot records from the voting terminals and judge's console. The very data that links a paper ballot to an electronic, encrypted ballot creates a potential vulnerability. Since some individual paper ballot summaries will be selected for post-election audit and made public at that time, we are careful to not include any data on the voter's take-hme receipt which can be associated with the corresonding paper balot summary.




\subsection{Absentee and Provisional Ballots}
There are several methods available for incorporating ballots which are not cast within the \projname system, such as absentee and provisional ballots.  The simplest approach is to completely segregate votes and tallies, but this has several disadvantages, including a reduction in voter privacy and much lower assurance of the accuracy of the combined tally.

It may be possible to eliminate all ``external'' votes by providing electronic means for capturing provisional and remote ballots.  However, for the initial design of the \projname system, we have chosen to avoid this complexity.  Instead, we ask that voting officials receive external votes and enter them into the \projname system as a proxy for voters.  While this still does not allow remote voters to audit their own ballots, the privacy-preserving risk-limiting audit step is still able to detect any substantive deviations between the paper records of external voters and their electronically recorded ballots.  This provides more supporting evidence of the veracity of the outcome without reducing voter privacy.

\section{Conclusions and Future Work}

\label{sec:future}
\label{sec:conclusion}

In many ways, \projname is a straightforward evolution from existing
commercial voting systems, like the Hart InterCivic eSlate, mixing in
advanced cryptography, software engineering, usability, and auditing
techniques from the research literature in a way that will go largely
unnoticed by most voters, while having a huge impact on the
reliability, accuracy, fraud-resistance, and transparency of
elections. Due to space constraints, this document doesn't even
scratch the surface of the many pragmatic features that our election
administration colleagues have specified based on their experience
running prior elections. Clearly, we're long overdue for election
systems engineered with all the knowledge we now have available.

\projname also opens the door to a variety of interesting future
directions. For example, while \projname is intended to service any
given county as an island unto itself, there's no reason why it cannot
also support {\em remote voting}, where ballot definitions could be
transmitted anywhere a voter wishes to vote, and results sent back
home. By virtue of \projname's cryptographic mechanisms, such a remote
vote is really no different than a local provisional vote and can be
resolved in a similar fashion, preserving the anonymity of the
voter. (A variation on this idea was earlier proposed as the RemoteBox
extension~\cite{remotebox08} to VoteBox~\cite{sandler08votebox}.)
This could have important ramifications for overseas and military
voters with access to a suitable impromptu polling place, e.g., on a
military base or in a consular office.

(We do not want to suggest that \projname
would be suitable for {\em Internet} voting. Using computers of
unknown provenance, with inevitable malware infections, and
without any systematic way to prevent voter bribery or coercion,
would be a foolhardy way to cast ballots.)

\projname anticipates the possibility that voting machine
hardware might be nothing more than commodity computers running custom
software. It remains unclear whether off-the-shelf computers can be
procured to satisfy all the requirements of voting systems (e.g.,
long-term storage without necessarily having any climate control, or
having enough battery life to last for a full day of usage), but
perhaps such configurations might be possible.


\bibliographystyle{acm}
\bibliography{votebox,security,policy,hfbib,iowa,auth,ucd-elect,shankar,arbaugh,mebane,smartphones,epstein,jhalderm,daw,stark,star}

\end{document}